\documentclass[%
 reprint,
 amsmath,amssymb,
 aps,
]{revtex4-1}

\usepackage{graphicx}
\usepackage{dcolumn}
\usepackage{bm}
\usepackage{natbib}
\usepackage{multirow}

\begin{document}

\preprint{APS/123-QED}

\title{Student Epistemological Framing on Paper-Based Assessments}

\author{Kelli Shar}
\affiliation{%
 Department of Chemistry, Biochemistry, and Physics, University of Tampa, Tampa, Florida 33606\\
}%


\author{Rosemary S. Russ}
\affiliation{
 Department of Curriculum and Instruction, University of Wisconsin,
 Madison, Wisconsin 53706\\
}%
\author{James T. Laverty}
\affiliation{%
 Department of Physics, Kansas State University, Manhattan, Kansas 66506\\
 \email{\textit{Correspondence to:} laverty@ksu.edu}
}%


\date{\today}

\begin{abstract}
 Assessments are usually thought of as ways for instructors to get information from students.  In this work, we flip this perspective and explore how assessments communicate information to students. Specifically, we consider how assessments may provide information about what faculty and/or researchers think it means to know and do physics, i.e. their epistemologies.  Using data from students completing assessment questions during one-on-one think aloud interviews, we explore how assessment features did (or did not) impact student engagement with the assessment problems. We analyze video recordings and transcripts to infer the epistemological framings and resources students use while completing introductory-level physics problems. Students' framings tended to be fairly stable, but when shifts occurred, they were triggered by a shift in epistemological resource, which can be activated by assessment feature. This work extends existing work on epistemological framing into the realm of assessment and allows us to consider the effects of assessments on our students' understanding of physics teaching and learning.
\end{abstract}

\pacs{Valid PACS appear here}

\maketitle


\section{\label{chapter:introduction}Introduction}
     As physics educators at undergraduate institutions, we are all aware of the importance of assessment in our classes. Individual assessments help us and our students understand whether students learned the content and skills we painstakingly taught them. Additionally, we assume that how a student performs on assessments throughout the semester helps us evaluate and track student progress. Students in introductory physics frequently take multiple, high-stakes assessments or exams each semester along with weekly homework assignments and lab reports. Assessment occurs often and in a variety of ways.
 
Consider a common assessment item on introductory physics exams - a numerical problem that requires calculation to solve. Imagine a question such as:

\begin{quote}
Diego is standing on a scale in an elevator and the elevator starts to accelerate upwards at 3 $ m/s^{2}$. If Diego weighs 71 kg on his scale at home, how much will the scale read while the elevator is accelerating?\\
\end{quote}

We suspect this type of question is familiar to many of our readers. We suspect the following solution is also familiar. 

\begin{quote}
$F = ma$\\
$F_{gravity} = ma_{elevator}$\\
$mg = ma_{elevator}$\\
$(71)(9.8) = m(3)$\\
$232 ~kg$\\
Units match!\\
\end{quote}

As instructors, we might find ourselves horrified by this student's solution. It is definitely wrong! They may have matched units but they entirely forgot the normal force. Many of us can relate to seeing such solutions on our exams and being confused that the students did not learn what we taught them. 

In these moments, our default is often to become frustrated - either with students for not learning enough content to solve the problem, or with ourselves for not teaching it well enough. That is, we attribute their lack of performance either to them or to our teaching. In this paper, we suggest that there may be a mechanism other than poor content knowledge that accounts for what looks like ``poor'' performance. 

Specifically, if we look closer, we can understand what the student is doing here as a type of pattern matching, or numerical plug 'n' chug using formulas she is familiar with\cite{tuminaro_elements_2007}. She is solving the problem using the symbols and mathematical formalisms we use in class and teach students to use. She has found the force of gravity and appropriately substituted in the acceleration of the elevator. In that way, she is solving the problem in a way that looks like other problems she has done.

Viewing student performance through this lens assumes that students are working to make sense of the assessment \textit{in the terms they think we want}. The logic is something like: ``I have seen my teacher use this formula in problems like this, so I will do the same thing here." This perspective on student engagement in assessment removes the assumption that students are being foolish when they do things that appear outlandish to experts.  Instead, it assumes they are trying to do what we have asked them - even taught them - to do.

In this paper, we explore this perspective to move away from our traditionally deficit-oriented model of assessment. Specifically, using the theoretical machinery of epistemological framing and resources, we seek to understand student performance as a reasonable - rather than misguided - interpretation of what assessment writers have asked them to do. We draw on data from think-aloud interviews with undergraduate students in introductory physics to (a) explore how students frame assessment, (b) unpack the knowledge resources that underlie those frames, and (c) document stabilities and dynamics in resources and frames within the context of assessment items.  Specifically, we describe how assessments may send students messages about what types of knowledge and knowledge production activities are appropriate to use in the assessment context. We discuss implications for these findings for educational research and assessment practice.

\section{\label{chapter:litreview}Literature Review}

\subsection{\label{sec:epist}Epistemology}

Within psychology and education, researchers have long been interested in how people understand the nature of knowledge and learning \cite{King_Kitchener_2002}. These understandings are referred to as a person's epistemology. A person's epistemology involves their knowledge and beliefs about the nature of knowledge itself and how knowledge is built and evaluated. Since the 1950s, scholars have proposed a variety of dimensions of epistemology. For example, epistemological knowledge includes knowledge about the goal or aims of knowledge construction \cite{berland_epistemological_2016}, how knowledge is structured \cite{chinn_expanding_2011}, how knowledge is justified\cite{Sandoval_2011}, and the appropriate activities to use in constructing knowledge \cite{Hammer_Elby_2002}. Scholars have explored personal epistemologies, which describe how people view their own knowledge and learning, and scientific epistemologies which describe how professional scientists construct knowledge and learning \cite{Russ_2014}. 

Much of the work understanding epistemology has taken place in educational settings \cite{Hofer_Pintrich_1997}. The focus on educational settings arises because of the particular importance of epistemology for learning. Specifically, there is the assumption that ``epistemological premises are a part of and an influence on the cognitive processes of thinking and learning” \cite{Hofer_Pintrich_1997}. In his cognitive model of learning physics specifically, Redish calls epistemology a control structure in that it ``interact[s] strongly with (and often controls) the [knowledge] resources students have for creating knowledge” (from p. 30)\cite{Redish_2004}. That is, a person's epistemology can impact the way they engage in learning by dictating the kinds of knowledge they use and the ways they use it.


Of particular importance for us in this work is the finding that what has often been labeled learning ``difficulties" \cite{Bollen_2017, Smith_Christensen_2015, Wilcox_Pollock_2015} may instead ``stem in part from [...] epistemology" \cite{Lising_Elby_2004}. For example, Lising and Elby present the case of a student Jan, who, despite possessing all the knowledge and skills needed to make sense of a physics tutorial, does not do so because her epistemology, or her understanding of what knowledge and knowledge building she is supposed to do in physics class, ``gets in the way'' (from p. 381)\cite{Lising_Elby_2004}. Epistemology mediates content knowledge in introductory physics.

Despite substantial interest in  epistemology in physics learning since Hammer's initial introduction of it \cite{hammer_introductory_1994}, we do not yet know of any who have directly examined epistemology in the context of assessment in undergraduate physics. The work has been constrained mostly to classroom or classroom-like contexts. In our work, we explore the hypothesis that the same mechanism at work in the case of Jan - the same ``epistemological effect'' \cite{Lising_Elby_2004} - might be at play in assessments. Specifically, we explore what epistemologies are active when students use their content knowledge to complete assessments.

\subsection{Assessment Design}

What types of assessments do our undergraduate students typically engage in? Assessment is a central pillar of our current education system and is often divided into two types: formative and summative.  Here, we will use the definition of formative assessments as ``intended to provide feedback to the system to inform next steps for learning'' and summative assessments as measures ``of individual achievement''\cite{council_knowing_2001}. Both of these definitions focus on assessments as ways to get information about what students know.  In this paper, we focus on summative assessments. Beyond these two definitions, current theories of summative assessment design focus explicitly on how we get information about students' current understanding.


The Assessment Triangle, described in {\it Knowing What Students Know}, has been central to assessment development for nearly twenty years.  It focuses on three interconnected models for developing assessments: cognition, interpretation, and observation\cite{council_knowing_2001}. Taken together, these three models (aka, the Assessment Triangle) are designed to treat summative assessment as an evidentiary argument, focused on designing tasks that allow the instructor or researcher to gather evidence to make claims about student knowledge.  

More recent examples of assessment design theories build on the Assessment Triangle model. Approaches such as Evidence-Centered Design \cite{Mislevy_2007, Mislevy_Haertel_2006, Mislevy_2002, Almond_steinberg_2002, Steinberg_Mislevy_2003} and the BEAR Assessment System \cite{Wilson_2005} also focus on obtaining evidence to support claims of student knowledge. Both of these approaches have been highlighted as promising ways to assess the Next Generation Science Standards\cite{ngss_lead_states_next_2013,council_developing_2013}.

The {\it Standards for Educational and Psychological Testing} states, ``Test development is the process of producing a measure of some aspect of an individual's knowledge...''\cite{aera_standards_2014}. In physics in particular, Adams and Wieman have argued that the development of concept inventories in PER typically follows the steps outlined by this document\cite{adams_development_2011}.

In each of these approaches to designing assessments, the focus is on obtaining information about student knowledge and little to no attention is paid to the messages that these assessments send to students. Assessments developed in these ways are assumed to be measurement instruments to get data about the knowledge of students in the same way a thermometer is a measurement instrument to get data about the temperature of a water bath; that is, that the measurement does not affect the system (or affects it minimally). Given this state of assessment design theories, we suggest that instructors and other individuals designing tests for physics courses probably do not think about the messages being sent to students either.

\subsection{Assessments in PER}

The history of PER includes the development of many standardized assessments.  As of this writing, the website Physport currently lists 93 research-based assessments, divided into 6 categories: Content knowledge (63), Problem-solving (2), Scientific reasoning (2), Lab skills (6), Beliefs / Attitudes (14), Interactive teaching (6)\cite{physport}. These assessments provide straightforward, off-the-shelf ways to evaluate student learning\cite{madsen_research-based_2016}. Because of this, they have been used to evaluate different learning environments, instructional strategies, and curricula (among other things)\cite{hake_interactive-engagement_1998,freeman_active_2014,von_korff_secondary_2016,brewe_toward_2010,kohlmyer_tale_2009,caballero_comparing_2012,madsen_gender_2013, brewe_toward_2010, lorenzo_reducing_2006}.

Within the assessment culture of undergraduate physics, most of the assessments are focused exclusively on evaluating student content knowledge. Further, many do so in a way that tacitly neglects students' understanding of either physics generally or the assessment in particular. There are a few assessments that attempt to directly measure student epistemologies. For example, the Redish, Saul, and Steinberg declare, ``[W]e describe the Maryland Physics Expectations survey; a 34-item Likert-scale  agree/disagree survey that probes student attitudes, beliefs, and assumptions about physics'' \cite{redish_student_1998} and Adams et al start their abstract, ``The Colorado Learning Attitudes about Science Survey (CLASS) is a new instrument designed to measure student beliefs about physics and about learning physics\cite{adams_new_2005}.'' However, we are not aware of any assessments that are designed to study or explicitly understand the connection between student epistemology and assessment of content knowledge.


\section{\label{chapter:theoretical}Theoretical Framework}

As described above, researchers interested in epistemology have not yet examined the ways in which student understandings of knowledge and learning are present in their engagement in assessments. Similarly, researchers interested in assessment overlooked the ways in which assessments are understood from the perspective of knowledge and learning. Here, we turn our analytic attention to bridging the gap between those two literatures. In what follows, we describe our theoretical stance on epistemology and its implications for our research on assessments.

\subsection{\label{sec:epistframing}Epistemological Framing}

Understanding the ways in which epistemology interacts with student engagement in assessment requires a precise conceptualization of epistemology. When it was first conceptualized, epistemology was understood as a set of categories that students adopt and then progress through over the course of their lifetime. These categories applied to all aspects of their learning at any given age.\cite{King_Kitchener_2002}

However, research shows significant ``flexibility and variability in student reasoning" in K-16 science and physics classrooms\cite{hammer_resources_2005}. As such, rather than a stable model of epistemology, we adopt a model of epistemology rooted in the sociolinguistic and anthropological construct of framing\cite{Machlachlan_Reid_1994}\cite{Tannen_1993}. Framing is a person's sense of ``What is it that is going on here?" In recent work in physics education, scholars describe epistemological framing \cite{scherr_student_2009} as students' answer to ``How should I approach knowledge?" \cite{hammer_resources_2005} Epistemological framing, then, is the tacit stance students take toward learning-based activities \cite{Odden_Russ_2018}.
    
A key feature of framing in general, and epistemological framing in particular, is that it is contextual and dynamic rather than stable across time and place\cite{tannen_interactive_1987}. For example, when a student enters a science class learning about electrostatics, she likely thinks very differently about knowledge than when she is in a discussion with her friends about what pizza to order. Even more, she is likely to think differently about learning during a portion of science class that is a lecture versus a small group discussion.\cite{Russ_Luna_2013} Even within small group discussion, students' sense of what knowledge and knowledge building activities should be used can shift dramatically\cite{scherr_student_2009}. This contextuality means that framing is highly dynamic. Because people shift their understandings of knowledge in different contexts, framing must also change over the time scale of contextual change (hours, minutes, and seconds) rather than over the scale of a lifetime.

Existing research on framing in undergraduate physics education suggests that students adopt a variety of frames when engaged in learning physics. For example, several scholars have explored the sensemaking frame in which students reason about physical phenomena by constructing an explanation and then filling in a gap in that explanation\cite{hammer_resources_2005, hutchison_attending_2010, Odden_Russ_2018}.

Bing and Redish\cite{bing_analyzing_2009} identify four common epistemological framings students adopt during physics problem solving: Calculation, physical mapping, invoking authority, and math consistency. In their work exploring framing in quantum mechanics, Modir, Thompson, and Sayre (2017) describe a set of frames that differ along two dimensions -- whether students draw on mathematics or physics, and whether they are engaged in algorithmic or conceptual thinking\cite{Modir_Bahar_2017}. Each of these framings involves different - though not necessarily better or worse - understandings of knowledge and knowledge construction in physics class.  

In addition to identifying framings that are prevalent in physics, scholars have also focused on whether and how students move between framings. Hammer and his colleagues describe transitions between mathematical manipulation and intuitive sensemaking that are both short-lived and lasting\cite{hammer_resources_2005}. Bing and Redish identified their four frames by explicitly looking for and unpacking shifts\cite{bing_analyzing_2009} and have suggested that the frequency and fluency at moving between framings is part of becoming and expert\cite{bing_epistemic_2012}. This work highlights the dynamic nature of epistemological framing. 

The theory of epistemological framing suggests the need to refine our question of interest even further. Specifically, this framework suggests that students may not adopt a single epistemology during assessment. Instead, they may transition between multiple framings. As such, we now ask: How do student epistemological framings influence their engagement in assessment tasks?
    
\subsection{\label{sec:eframes}Epistemological Resources}

As we have described, existing scholarship within physics education highlights the dynamics of framing. To examine and document those dynamics, much of the literature focused on identifying observable shifts in behaviors. Scherr and Hammer (2009) pioneered focus on behaviors with their careful analysis of students' verbal, non-verbal, and para-verbal behaviors in small group interactions\cite{scherr_student_2009}. They describe how ``different behavioral clusters are evidence of - and in dynamic interaction with - student epistemologies" (p. 148). This finding led to an explosion of work in PER that identified behavioral clusters and their associated framings.

For example, consider more closely the work of Modir, Thompson, and Sayre who identified four epistemological framings in upper level student problem solving\cite{Modir_Bahar_2017}. To identify framings, they describe how they ``reflected on [...] episodes [of student activity], seeking to answer `what's going on?' for each of them [...] we sought to capture changes in students’ discussion or behavior that might indicate a shift in the students’ problem solving processes" (p. 020108). Here, they focus on the behaviors associated with their framing.
   
This focus on behavior makes sense and has given the field substantial traction in identifying moments when epistemological framing shapes student engagement and learning. However, a feature of the theory of epistemological framing that is commonly left out of the literature in physics education is that framings are local collections of epistemological resources \cite{hammer_resources_2005, Rosenberg_Rock_Cycle_2006}. Specifically, our model of epistemology - grounded in the work of Hammer and Elby\cite{Hammer_Elby_2002} - assumes that rather than being unitary entities that stably exist as a coherent unit, framings are more like networks of many smaller elements that are all activated together in context\cite{hammer_resources_2005}.
   
The small elements that make up epistemological frames (or what Rosenberg, Hammer, and Phelan call ``multiple local coherences") are epistemological resources\cite{Rosenberg_Rock_Cycle_2006}. These resources, which are finer-grained elements of knowledge, are assembled anew in each context to create a person's in-the-moment epistemology. This finer-grained structure affords contextuality in a person's epistemology.
   
A number of potential epistemological resources have been proposed that govern student behavior in knowledge-production contexts. Some deal with the nature of knowledge itself:  the form of the knowledge product\cite{collins_epistemic_1993} or the source of the knowledge\cite{Hammer_Elby_2002}. Others deal with how knowledge is produced or constructed: knowledge production activities\cite{Russ_Luna_2013} or the goals of those activities\cite{berland_epistemological_2016}. 

Our model, as drawn from Hammer, Elby, Redish, and Scherr, for the relationship between epistemological resources and frames is that frames are local coherences of epistemological resources\cite{hammer_resources_2005}. This means that, when we identify frames by behaviors, we can expect to find a fairly stable set of resources associated with that frame. Alternatively, if we see a group of resources reliably coming up together, we can expect there to be a set of associated behaviors, or frames, each time. Some resources are likely to be more central to the frame\cite{Rosenberg_Rock_Cycle_2006}. Others may be reliant on those more central resources. In our work, we focus specifically on the resources other empirical work has found relevant in learning science and physics \cite{Russ_Luna_2013, Rosenberg_Rock_Cycle_2006}.
  
In our work, we hypothesize that understanding the specific resources that make up the framings will give us more insight into how and why students engage in learning in particular ways than merely identifying their epistemological framings as a whole. This assumption follows the work of Rosenberg, Hammer, and Phelan (2006) and Russ and Luna (2013) who each go an analytic level below framings to the level of resources\cite{Rosenberg_Rock_Cycle_2006, Russ_Luna_2013}. Specifically, We apply this assumption to the context of assessments and ask: How are the epistemological resources that make up student epistemological framings evident in their (potentially dynamic) engagement in assessment tasks?

\section{\label{chapter:rqs}Research Questions}
    Our everyday experience as physics instructors leads us to question why students engage with assessments in the way they do. Further, our underlying asset-oriented perspective toward student learning leads us to reject the existing deficit-based explanations in the literature\cite{scherr_modeling_2007}. As such, we began this research with the question: Why might students reasonably engage with assessments in the way that they do?

As a result of our theoretical framework, we refine and extend our question:

\begin{enumerate}
\item What epistemological resources make up the framings students adopt during assessments?
\item How do these epistemological resources influence their dynamic engagement in assessment tasks?
\end{enumerate}

Specifically, this work seeks to understand whether existing analyses and assumptions of how students dynamically bring different forms of knowledge into physics learning can be extended into the realm of assessment.
    
\section{\label{chapter:methods}Methods}


\subsection{\label{sec:subjectselection}Participants}

In this article, we analyze data collected as part of a different study on how to  assess scientific practices in introductory physics courses \cite{national_research_council_framework_2012}. The participants in this study were predominantly engineering majors enrolled in the first or second semester of a calculus-based, introductory level physics course. The interviews were conducted near the end of the semester. Students volunteered to take part in the study and were compensated for their time (equivalent to \$20). No preference was given to their performance in the course. Pseudonyms are used throughout the paper. The study collected data from two groups of ten students, with each group of ten receiving a separate exam.

\subsection{\label{sec:datacollection}Data Collection}

The data was collected as individual think-aloud interviews designed to replicate the context of a summative assessment. To do so, interviewers printed each list of questions as a formatted test, had subjects write their names at the top, answered only clarifying questions, and told students to complete the questions ``as a test" with no permission needed to move on to another question. Each interview took each student 45 to 90 minutes to complete. The students had unlimited time to complete the questions.

The questions on the tests were developed using the 3D-LAP to elicit evidence of students abilities to engage in the scientific practices\cite{laverty_characterizing_2016}. One of the exams focused on the scientific practice of Using Mathematics and the other focused on Developing and Using Models. Students solved physics problems on a variety of first-semester Physics subjects. The exam included both selected and constructed response items. Only the constructed response items were used in this analysis.


\subsection{Data Selection}
        
From the recorded interviews, there were two sets of 10 students. One student from each group of 10 did not have usable audio, resulting in 9 recorded interviews for both problem sets. The assessment given to the first group of 9 students featured 3 constructed response questions. The assessment given to the second group of 9 students featured 5 constructed response questions. This yields a total of 72 instances of students answering a constructed response question. Of these 72 responses, 5 were not used. Reasons for not using a response included the student not attempting that specific problem or stating that they have seen the problem before and know the answer (and, thus, do not engage in the problem solving process). Below is an excerpt from Mark.
    
    \begin{quote}
        \textbf{Mark}: Alright, so we had this exact problem on our test last year. 
    \end{quote}
    
Here, Mark states he knew the answer and goes on to repeat the answer he remembered from his exam. Because we were specifically interested in how students engaged with the features of this specific assessment question, we decided not to use his response or similar responses where students do not engage in problem solving.

\begin{figure*}
    \centering
    \includegraphics{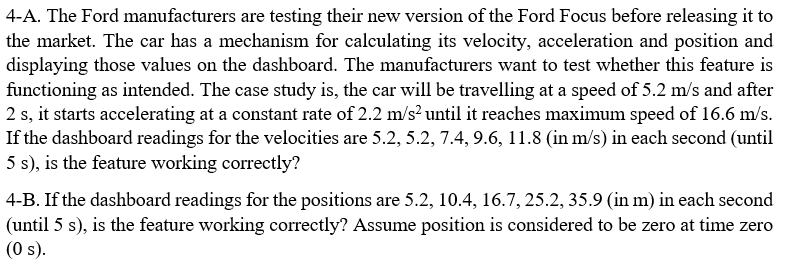}
    \caption{The Car Problem. }
    \label{fig:Ford_Focus}
\end{figure*}

\begin{figure*}
    \centering
    \includegraphics{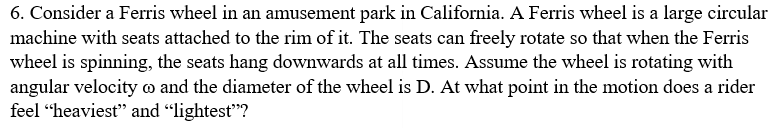}
    \caption{The Ferris Wheel Problem.}
    \label{fig:Ferris_Wheel}
\end{figure*}

\begin{figure*}
    \centering
    \includegraphics{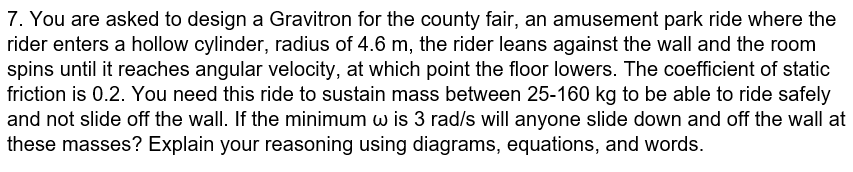}
    \caption{The Gravitron Problem}
    \label{fig:Gravitron}
\end{figure*}

The examples we use throughout the paper come from specific assessment questions we will refer to as the Car problem and the Ferris Wheel problem, shown in Figures \ref{fig:Ford_Focus} and \ref{fig:Ferris_Wheel} respectively. We also transcribed the audio from what we will refer to as the Gravitron problem, which we chose as an example of students ardently staying in their problem solving strategy. The full Gravitron question is shown in Figure \ref{fig:Gravitron}. Although we largely reference those problems in the cases we present, all constructed response questions were analyzed via the process described in the Data Analysis section. 

\subsection{\label{sec:analysis}Data Analysis}



After reducing our data set to only instances where students engaged in problem solving with usable audio, we were left with 67 attempts at solving a physics problem. Each instance corresponds to a specific student solving a specific problem. Of the remaining 67 instances, all were coded for epistemological frames. A sample of 20 instances were transcribed and coded for epistemological resources. After coding each full sentence expressed by the subject, we then compared when frame shifts occurred to when shifts in individual resource shifts occurred. In this section, we describe how we coded the data and offer an example of coding one instance.

\subsubsection{Coding Epistemological Frames}

For this particular project, we opted to use epistemological frames that had already been identified in the literature. In particular, We wanted to use frames that could be identified by looking at behaviors, a process originally described by Scherr and Hammer\cite{scherr_student_2009}. We chose the frames described in Chari et al.\cite{Chari_frames_2017}. We selected this set of frames and behaviors primarily because they were already identified and because they mapped well onto our data. Chari et al. used their frames to analyze group problem solving interviews, but the language used to define the frames does not limit them to group or individual work. Below are the characteristics and behaviors associated with each fram as described in Chari et al. (2017)\cite{Chari_frames_2017}.
        
    \begin{description}
        \item[Conceptual Physics] Students and instructors are in this frame when they discuss physics scenarios and phenomenon, about properties of physics quantities related to the task at hand. They may also exploit the symmetry of a physical system by investigating related concepts.
        \item[Algorithmic Physics] In the Algorithmic Physics frame, students recall physics equations or apply physics knowledge to re-arrange known equations using math. Students may also derive expressions for specific cases from a general physics equation or validate an expression via dimensional analysis.
        \item[Algorithmic Math]This frame refers to performing mathematical computation by following well-established protocols without questioning the validity of those protocols e.g. solving an equation or computing an integral.
        \item[Conceptual Math]Students are in this frame when they exploit properties of mathematical constructs to quickly obtain a result without diving into algorithmic manipulation e.g. noticing that all the odd terms in a sun are equal to zero.
    \end{description}

Using these frames and definitions, we analyzed the recorded video interviews using a top-down coding scheme to identify frames and noted each time a student switched from one set of behaviors to another and identified this as a frame shift.

\subsubsection{Coding Epistemological Resources}

As we were specifically interested in what causes a frame shift, we transcribed 20 student solutions: the three examples we found of a clear frame shift, and all responses to two questions where almost all students solved the problem in the same frame. These were coded for resources to investigate what prompted students to stay in this frame.


Next, we coded all transcribed attempts for epistemological resources. Each full sentence of the transcribed interview was sectioned, and we inferred epistemic resources according to definitions given in Table \ref{tab:resourcedef}. We grouped by each full sentence because that was as large as we could go while still determining a resource shift and as small as we could go while still having the context to determine a resource.

To code the resources, We used categories defined in Hammer and Elby's 2002 paper \textit{On the form of a personal epistemology}\cite{Hammer_Elby_2002}. We specifically focused on Nature of Knowledge, Source of Knowledge, Epistemic Activity, and Epistemic Source because we found clear indicators of these in the words and actions of the students. 

The individual resources we used also came from Hammer and Elby (top-down coding).  In addition to re-purposing codes from the literature on epistemology, we also generated codes from the data in an emergent fashion (bottom-up coding)\cite{coffey_qualitative_1996}. These codes are Equation Sheet as a Knowledge Source; and Number or Reasoning as Epistemic Forms. The working definitions and an example from the data for each resource are listed in Table \ref{tab:resourcedef}. Only resources that appeared in the data are listed.

We did not limit Epistemic Form to the answer type requested by the assessment question. Instead, we decided there could be different epistemic forms possible for one question and for different portions of the problem-solving process. For example, suppose a multiple choice question asks students to find an acceleration, given a mass and net force, and lists four different values for acceleration. To solve, students could plug values into Newton's Second Law and choose the acceleration closest to their answer. They could also reason through the options listed and pick the most probable value and look for the answer with the correct units. We believe these two students, although their ultimate answers were a letter, had a difference in Epistemic Form because they were working towards different ending conditions, a number or reasoning, that they later translated into a letter.

\begin{table*}[htb]
\begin{center}
\begin{tabular}{p{3.2cm}|>{\raggedright\arraybackslash}p{3.2cm}|p{10.8cm}}

Resource Type & Resource Definitions & Example from Data \\ 
\hline

\multirow{4}{3.2cm}{\textbf{Knowledge Type} \newline If you asked the student how they know this, they will say...} & \textbf{Propagated} \newline... because someone told me. & Okay so centripetal force is going to be equal to r w squared \\
\cline{2-3}
 & \textbf{Fabricated}\newline...because I figured it out using my prior knowledge. & Your potential energy is going to be the highest when you're at the top, lowest at the bottom, so I think those are going to be the points at which you're going to be feeling heaviest \\
\cline{2-3}
 & \textbf{Directly Perceived} \newline...because it is apparent or obvious. & The radius is 4.6 \\
\cline{2-3}
 & \textbf{Intuitive} \newline...because I feel like this is true. & Um, like any equation doesn't seem to give me any intuition \\
\hline

\multirow{4}{3.2cm}{\textbf{Knowledge Source} \newline If you asked the student where they got this knowledge from, they would say...} & \textbf{Equation Sheet} \newline...from my equation sheet & Okay, so, we look at our rotational kinematics equations \\
\cline{2-3}
 & \textbf{Physical Experience} \newline...from my memory of an experience I had or observed & So that's based off my life experiences, not any equations \\
\cline{2-3}
 & \textbf{Authority}\newline...from something my professor or textbook told me. & I'm trying to recall what the, uh, like a similar problem we did \\
\cline{2-3}
 & \textbf{Memory} \newline...from my memory of solving another problem. & So I know there will be a friction force which is going to be equal to m g mu \\
\hline

\multirow{4}{3.2cm}{\textbf{Epistemic Activity} \newline The student obtained this knowledge by...} & \textbf{Accumulating} \newline...gathering information from external sources. & So this is just off to the side, I just kind of like to write down some of the stuff for later use\\
\cline{2-3}
 & \textbf{Forming} \newline ...creating new knowledge about the specific problem from knowledge they already had. & I guess all that matters is the direction of the force \\
\cline{2-3}
 & \textbf{Checking}\newline...reconciling their thoughts with another piece of information. & If the person weighs more, it's going to be easier for them to drop straight down and harder for them to go... Is that right? (Checks equation sheet) \\
\cline{2-3}
 & \textbf{Causal Story-telling}\newline...attributing a cause to each effect. & Alright, so this force has to keep them from falling down \\
\hline

\multirow{4}{3.2cm}{\textbf{Epistemic Form} \newline The student believes the answer to their question will be indicated by...} & \textbf{Number} \newline ...a quantity. & Okay, we need to find alpha \\
\cline{2-3}
 & \textbf{Reasoning} \newline...an application of conceptual knowledge. & It doesn't tell you the mass. So, I guess because this is a conceptual question \\

\end{tabular}
\caption{Definitions used for each Epistemological Resource}
       \label{tab:resourcedef}
\end{center}
\end{table*}

\subsubsection{Inter-rater Reliability}
The first author coded all 67 transcripts for their frames and frame shifts. The second author again coded a subset (8) of the transcripts. The two coders agreed on all codes (100\%), yielding perfect agreement. 

Additionally, the first author coded each full sentence of all 67 transcripts for each of the four dimensions of epistemological resources. To check the reliability of this coding, the second author coded a 10\% subset of the transcripts (7). We then calculated percent agreement across the doubly-coded transcripts for each dimension. The percent agreement for each epistemological dimension (aka coding category) was greater than or equal to 80\%. Specifically, agreement was 81\%, 84\%, 81\%, and 98\% for Knowledge Type, Knowledge Source, Epistemic Activity, and Epistemic Form, respectively, which is considered acceptable agreement for qualitative coding in research\cite{IRR_2012}.



\subsubsection{Example Analysis}
Here we include an example of how this analysis is done in practice. Figure \ref{fig:Gravitron}, or ``The Gravitron Problem" features a problem that gives students the radius, coefficient of friction, and angular velocity of a Gravitron ride and asks them to determine if the ride is safe for riders within a range of masses. Spoiler alert: The safety of the rider is not dependent on their mass. Below is a transcript of Erica beginning to solve the problem:
\begin{quote}
    \textbf{Erica:} Okay, so I’m going to get the rotational ones (picks up equation sheet, starts copying equations onto paper). Um, so you just want to find like your minimum 25 kilograms and your maximum 160 kilograms and then any answer you get between that is that going to, um, stay on the wall. (Looks at problem) Um, 3 radians a second, so I’m just going to write that up here (writing on paper) … radians per second… (typing into calculator) Let me get that number… 60 seconds for one minute and it’s 2 pi radians for one revolution (types into calculator) or something like that.
\end{quote}

First, we used behaviors to determine the Epistemic Frame each student was working in. Erica starts the problem by rearranging known physics equations that she can use to input given values. She believes the output of her function will tell her whether some people will slide off the Gravitron. For these reasons, we determined that Erica began the problem in an Algorithmic Physics frame.

Second, we grouped coherent thoughts and coded each thought using the four categories of Epistemic Resources defined in Table \ref{tab:resourcedef}. Erica's first full sentence is ``Okay, so I'm going to get the rotational ones." We coded her Type of Knowledge as Propagated because the knowledge of the physics equations is communicated directly to Erica through the equation sheet. We coded her source of knowledge as her equation sheet. We coded her Epistemic Activity as Accumulating because Erica is gathering knowledge from an external source. Erica's epistemic source is ambiguous here because she could be intending to use the equation to consider how the different quantities given affect each other and reason her way to an answer. We used additional context from later in the problem to determine that Erica intended to use the equation to input her given values and output a number that would indicate which riders would be safe, so we coded her Epistemic Form as Number.

We repeated this process across all 20 solution attempts that we transcribed. 

\section{\label{chapter:results}Results}
    \subsection{\label{sec:definedby}Different Frames are Made up of Different Sets of Resources}

Existing literature examining student epistemologies describes how frames are made up of resources\cite{hammer_resources_2005}. Specifically, frames are ``local coherences" of resources\cite{Rosenberg_Rock_Cycle_2006}; that is to say groups of several different resources tend to co-occur and change together. For example, Russ and Luna (2013) identified distinct sets of resources one teacher drew on when she engaged in behaviors from different framings \cite{Russ_Luna_2013}. In one frame, she engaged in particular epistemological activities with particular epistemological goals and in another frame she used different activities and goals. 

Our independent coding of frames by behaviors and resources by utterance allows us to look for local coherences of resources within and across frames identified by other researchers\cite{Chari_frames_2017}. That is, we can engage in an analysis similar to that done by Russ and Luna (2013) and Rosenberg, Hammer, and Phelan (2006)\cite{Russ_Luna_2013, Rosenberg_Rock_Cycle_2006}. Although this result is not theoretically ``new" (by definition frames are made up of resources), we begin with it here for two reasons. First, other researchers in Physics Education Research have yet to unpack frames in terms of their underlying resources. This result demonstrates the feasibility of that work. Second, knowing the resources underlying the frames for assessment is essential for arguments we will make later in the results section.

We begin by looking at the framings themselves. Of the 67 responses used, three students' behaviors indicated they started in one frame and transitioned to another. These students are featured as case studies in Sections C and D. Of these three students, two began a problem in an Algorithmic Physics frame and transitioned to a Conceptual Physics frame, and one student started a problem in a Conceptual Physics frame and transitioned to an Algorithmic Physics frame. This yields a total of 70 frame instances observed.
    
Table \ref{tab:frameswresources} shows the resources we identified in each of the four frames. Within each frame, we identified a subset of resources often used when students were working in each frame. Across frames, different combinations of resources are used. Particular frames are associated with particular groups of resources different from other frames. This provides empiracle evidence for our claim as predicted from our theoretical perspective. 
  
\begin{table*}[thb]
\begin{tabular}{p{3.2cm}|p{3.2cm}|p{3.2cm}|p{3.2cm}|p{3.2cm}}

Epistemological & \multicolumn{4}{c}{Epistemological Resources} \\ 
\cline{2-5}

Frames N=70 & Knowledge Source & Knowledge Nature  & Epistemic Activity & Epistemic Form \\ \hline

Conceptual Physics \newline N=13 & Self \newline Physics Concepts & Fabricated & Forming \newline Comparing \newline Storytelling & Reasoning \\[1cm]

Algorithmic Physics \newline N=55 & Equation Sheet & Propagated & Accumulating & Number \\[1cm]

Algorithmic Math \newline N=1 & Math Protocol & Propagated & Accumulating \newline Computing & Number \\[1cm]

Conceptual Math \newline N=1 & Math Concepts & Fabricated & Comparing \newline Forming & Reasoning \\

\end{tabular}
\caption{Total instances of each frame observed and the resources associated with them.}
       \label{tab:frameswresources}
\end{table*}

\subsection{\label{sec:shifting}Students Seldom Shift Frame}

The introductory students in this study almost always began problems in an Algorithmic Physics frame. Looking across the 67 total initial frame instances, 54 students started in the Algorithmic Physics frame, 11 students started in the Conceptual Physics frame, one student started in an Algorithmic Math frame, and one student started in a Conceptual Math frame. 

For some questions, beginning and staying in the Algorithmic frame is an effective way to solve the problem. An example of this would be the Car problem that asked students to verify odometer readings at several times for a car. Students were given the car's acceleration and initial velocity. See Figure \ref{fig:Ford_Focus} for the full problem. Below is a quote from Lisa as she solves the problem:

\begin{quote}
    \textbf{Lisa:} So we can use this other equation if we want to confirm positions. So our V naught is going to be 5 point 2, 10 point 4, 16 point 7, 25 point 2, 35 point 9. So, using the equation to double check for position... (On her paper, Lisa isolates d, then types numbers into calculator) 16.7. Should I do it again? It might be fun. (Types into calculator) The next one is 25.2, which is as far as I want to check.
\end{quote}

Lisa finishes by indicating on her paper that the feature is working correctly. It was determined that Lisa was working in the Algorithmic Physics frame because Lisa first accumulates information given in the question and equation sheet, then manipulates a physics equation to isolate the desired variable, then inputs her given values into her new equation and obtains a number that she interprets as an answer.

In this case, the Algorithmic Physics frame serves Lisa well, and she has no need to transition. Other questions are more easily solved in a conceptual frame. Ideally, students could switch fluidly when they get “stuck” on a problem in their current frame, as experts do\cite{bing_epistemic_2012}.
However, our data shows a fairly stable initial frame, meaning that students seldom switch out of their initial frame after beginning the problem. We found the Algorithmic Physics frame implements a specific set of resources, as shown in Table \ref{tab:frameswresources}.

The question that showcases this best from the problem set is the Gravitron problem (Figure \ref{fig:Gravitron}). Students were specifically asked to show their reasoning using diagrams, equations, and words. The problem may have been solved easily in a Conceptual Physics frame by noticing that the force of friction and the force of gravity should be equal so as not to let the rider slide down, which results in mass cancelling out. The resulting equation shows that gravity will overcome the force of friction and all riders will fall down. Here is a quote from Amanda as she works to complete the Gravitron problem:
 
\begin{quote}
 
\textbf{Amanda:} So, right now I'm thinking about what to do for the Torque, to try to find or use some of the kinematic equations. I'm trying to remember how to do that. Find the alpha, but... So I know there will be a friction force (Draws or writes on paper) which is just going to be mg mu, and then... and that will keep you put at the wall minimum. So then, (Writes on paper) you can find torques for the minimum and the maximum mass. (Types into calculator, writes on paper) Okay, so now we have two torques. Hmm. I'm not sure where to go from there.
     
\end{quote}

Amanda begins the problem by attempting to re-arrange given rotational kinematics equations such that she could plug her given values into an equation and get an answer. For those reasons, we determined she was in the Algorithmic Physics frame.

Of the nine students who attempted the Gravitron problem, we observed all students beginning the problem in the Algorithmic Physics frame as indicated by their listed behaviors. Overall, students found the problem very challenging in this frame. Most did not solve the problem, but no students shifted their initial frame. 

Across problems, our data shows that students often started in and did not shift from the Algorithmic Physics frame. Analysis of students' statements in the Gravitron problem shows that students did not attempt another frame even when making little or no progress in their current frame. We provide possible reasons for this in the Discussion.

\subsection{\label{sec:resourcekicking}Kicking a Resource Can Shift a Frame}    
While students seldom shifted frames, we observed three instances of a frame shift. All were caused by a change in a single epistemological resource. This was the result of two main influences: interviewer intervention and assessment feature.

Our first example of this comes from a student named Luke working on the Car problem (Figure \ref{fig:Ford_Focus}). Below is the transcript of Luke beginning the problem:

        \begin{quote}
            \textbf{Luke:} Okay, so the positions are... these make sense. So, as its moving, it's accelerating at 2.2 meters per second every second. It's accelerating. So... 5.2... Since it's not accelerating during this time. Well, I guess I proved here that the difference in between is one second, so I guess that is the correct position. I feel like there is a simpler way of doing this that I'm overlooking.  Um... (re-reads the problem silently)
        \end{quote}

This segment spans two minutes. Luke begins the problem by thinking conceptually about acceleration and the quantities he was given. Notice how he does not start any mathematical calculation or manipulation, but rather expects to obtain an answer by reasoning through the problem. The interviewer, noticing Luke was on the wrong path, intervenes:

    \begin{quote}
        \textbf{Interviewer:} When you find the positions, you can compare. The question is giving position. It is asking for validation.\\
        \\
        \textbf{Luke:} Oh, I think I assumed the wrong thing when I did it this way. Because I assumed that... Well, I guess... So I assumed that... yeah. So, assuming that how fast it's going, starting at that velocity, I found that it would take one second to get from this point to this point, which verifies that this is correct. So then going to this point, I guess I would have to do this same thing again. That's just a lot of math.
    \end{quote}
   
This segment spanned one minute. After the intervention by the interviewer, Luke realizes that he can solve the problem by substituting his givens into a known physics equation, which is quite the relief. Initially, Luke attempts to solve the problem by contemplating the properties of the quantity acceleration he was given in the question. After the intervention, Luke finishes the question by substituting the given quantities into a physics equation on his equation sheet. We interpret this, first, to be a shift from a Conceptual Physics frame to an Algorithmic Physics frame after the intervention by the interviewer.

A closer look at the resources Luke uses reveals a shift in Epistemic Form at that moment. Before the intervention, Luke believes he can reason his way to an answer to the Yes or No question of whether the odometer is functioning properly on the car. We interpret his Epistemic Form at that moment to be Reasoning. During the intervention, Luke realizes that he can easily calculate a number that will reveal the answer to the Yes/No question. We argue that the intervention at the level of Epistemic Form is what caused Luke to shift from a Conceptual Physics to an Algorithmic Physics frame, and thus a new set of resources.

\subsection{\label{sec:asskicking}Assessment Features Can Shift a Resource, Which Shifts the Frame}
    
Outside of intervention, instructors may be able to influence their students' resource use, and thus their framing, through features of an assessment question. 

To show this, we have two examples of a student (Lisa and Jack) completing a problem where they were asked to determine where a rider feels heaviest and lightest on a Ferris Wheel. See Figure \ref{fig:Ferris_Wheel} for full problem. Students could solve the question in a Conceptual Physics frame by thinking about which direction the Normal Force points at different locations on the Ferris Wheel. Notice that the question does not give the students any quantities, and the variables given in terms of letters are constant. As you will read, this prevents students from successfully solving the problem in the Algorithmic Physics frame.

A total of nine students attempted this problem. Seven attempted the problem in what we determined to be a Conceptual Physics frame. Although we cannot conclude what made the students treat the Ferris Wheel problem differently, some admitted to having seen the problem before, so it is possible they already knew it could be solved conceptually. The remaining two students, Jack and Lisa, treated the problem very differently. First, let us look at how Lisa solves the Ferris Wheel question. Below is the transcript of Lisa starting the problem:

        \begin{quote}
            \textbf{Lisa:} Um... (picks up equation sheet) Where's centripetal force? (Writes down centripetal force equation) Hmm... (looks at equation sheet) not given any numbers. It's just weird to me because it seems like nothing is changing. I mean, this is just going to be D over 2. The mass of the person isn't changing. Angular velocity is not changing.
        \end{quote}

Lisa begins the problem by attempting to manipulate equations from her equation sheet and plug in known quantities that she can translate into a position on the Ferris Wheel as her final answer. For these reasons, we determined that Lisa starts in the Algorithmic Physics frame. She becomes frustrated when she realizes that this method will not yield an answer for her. After pausing, she changes her approach:

        \begin{quote}
            \textbf{Lisa:} All I'm thinking about is that your potential energy is going to be the highest when you're at the top, lowest at the bottom, so I think those are going to be the points at which you're going to be feeling heaviest, and those are the points... I don't know why I'm thinking this but I just keep thinking of a clock and a pendulum swinging or anyone being on a swing and your... the points where you feel like you're accelerating the fastest are the ones where you're crossing this vertical axis. But I couldn't tell you why.
        \end{quote}

After realizing that she cannot answer the problem by manipulating given physics equations, Lisa changes her approach by contemplating her physical experience of being on a swing and applying her knowledge of potential energy. For these reasons, we determined that Lisa switches to the Conceptual Physics frame. 

Initially, Lisa believes she can solve mathematically for an expression which she can translate into a position on a Ferris Wheel as her final answer. We interpret Lisa's epistemological form here as Number. Not providing numbers or varying quantities is a feature of the assessment that interrupts Lisa's initial epistemological form, thus forcing her to try something new to finish the question.

Next, Lisa believes she can determine the answer by reasoning through it, so we interpret her epistemic form to be Reasoning. This is a clear case of an assessment feature shifting an epistemic resource, thus shifting Lisa's epistemic frame.

We saw a very similar pattern in how Jack completed the problem as well. Below is the transcript of Jack starting the problem:

        \begin{quote}
            \textbf{Jack:} So this... for you to have this (points upward) that means you have to have a greater force acting on you because, like, which is like, you have to have a greater force acting on you because like... you'll have a greater force acting on you, and here you'll have a smaller force acting on you. 
            (Picks up equation sheet) We'll see, the angular velocity equation. I should look for the angular velocity equation. I should, yeah. (Flips through equation sheet) It's this one. Moment of inertia. Um, it doesn't tell you the mass. 
        \end{quote}

Jack begins the problem by thinking about which forces are acting on the rider at different points on the Ferris Wheel. He concludes that he needs the angular velocity equation to solve and finds it on his equation sheet. Because Jack attempted to solve the problem by plugging values into known physics equations, we conclude that he was initially in an Algorithmic Physics frame. Because Jack thought plugging values into the angular velocity equation should give him an answer he could translate into a position on a Ferris Wheel, we interpret his initial epistemic form to be a Number. Jack hits a roadblock when he realizes he cannot solve the problem this way, and decides to change his strategy:

        \begin{quote}
            \textbf{Jack:} So, I guess because this is a conceptual equation... I am going to guess, for that, um, okay, so the greatest force you would be feeling (points upward) since you're going up, that means you would feel the lightest at the top, and the heaviest at the bottom. Because, like, you're going, accelerating faster at the bottom, and your mass is the same, so like the force would be greater at the bottom than it would be at the top. Because you're accelerating faster at the bottom than at the top. So, yeah, that's my answer.
        \end{quote}

Jack finishes the problem by applying what he knows about the physics phenomenon of force. We determined Jack finished the problem in a Conceptual Physics frame. Jack also switched from trying to obtain a mathematical expression as his answer to trying to reason his way to an answer, so we determined his final epistemic form to be Reasoning. As Jack expresses, the assessment question not including numbers causes him believe the question is conceptual in nature, which causes him to change what he is doing by switching from plugging numbers into an equation to thinking about the direction in which several forces are pointing. We interpret this as the assessment feature (no numbers) shifting his epistemic form, which shifts his epistemic frame.


\section{\label{chapter:discussion}Discussion}

Our first claim is that Epistemological Frames (as identified by behaviors) can be defined by the Epistemological Resources used. Our analysis is consistent with much of the other work on problem solving. This suggests we can make sense of students' problem solving work during exams in much the same way we can make sense of their reasoning in other situations.

When students engage in problem solving during an assessment, they draw on a range of epistemological resources to do so. In Rosenberg et al.'s paper, \textit{Multiple Epistemological Coherences in an Eight Grade Discussion of the Rock Cycle}\cite{Rosenberg_Rock_Cycle_2006}, researchers noticed groups of epistemological resources that consistently appeared together when students engaged in specific problem solving strategies that they called ``epistemological coherences." 
They described that each group of resources was stable and reinforced the others. In our interpretation of this data, we understand their ``epistemological coherences'' to be epistemological frames as defined by student behaviors.

Additionally, defining epistemological frames by epistemological resources makes some intuitive sense. For example, ``Causal storytelling'' as an activity would not make sense in the Algorithmic Math frame where students are computing and executing well established protocol. Alternatively, one could imagine that students displaying the behavior associated with the Algorithmic Math frame must engage in Accumulating as they gather given information for their computation. Another example is that ``Number'' as an epistemological form would not make sense in a Conceptual Physics frame where students are thinking about physics phenomenon and reasoning their way to an answer, whereas ``Reasoning'' certainly would make sense.

Defining Frames in terms of resources also changes the way we think about their relationship to one another. This could provide a useful tool to researchers. For any given coherent thought given by a student, it may prove easier at times to determine three to four epistemological resources that can be matched with a particular frame than attempt to interpret sometimes ambiguous student behavior. Defining frames in terms of both resources and behaviors gives researchers on small teams an efficient way to check their frame coding for accuracy and consistency.

Our second claim is that novice physics students seldom shift frames.  Our results agree with previous findings. As discussed in the Literature Review, Bing \& Redish found that one marker of expertise in physics is the ability to switch between frames as necessary\cite{bing_epistemic_2012}. Ideally, students could shift frames when one fails to help them complete the task at hand. Students did not shift frames when their current frame proved unhelpful, such as in the specific example of the Gravitron. As discussed in Subsection B of the Results section, the Gravitron problem could have been solved very quickly in a Conceptual Physics frame, yet most students stayed in an Algorithmic Physics frame and struggled to arrive at an answer. 

Rather than believing students perform poorly at switching frames, we believe students may be consciously choosing to begin and stay in the Algorithmic Physics frame. This could be due to some Physics textbooks which specifically call for students to solve physics problems algorithmically\cite{knight_physics_2007,bauer_university_2013,mazur_principles_2014}. This could also be due to the course instructors who sometimes teach students an algorithm for solving physics problems \cite{wright_williams_1986, Heller_Hollabaugh_1992}. It is possible that some part of the assessment itself cues students to begin a problem in a particular frame. Students could also begin problems in an Algorithmic Physics frame because that is what they generally do in physics assessment. Our data does not reveal why students start in the Algorithmic Physics frame. This idea is discussed further below.

Our third and fourth claims are that Epistemological Frame shifts happen on the level of an individual Epistemological Resource and that assessment features can prompt that resource shift. Our data implies that researchers could focus on how to shift resources as the means to the desired ends of frame shifting. We did not observe simply telling students to solve the problem conceptually or algorithmically to be effective in shifting frames. Notice that in the Gravitron problem (Figure \ref{fig:Gravitron}) students are explicitly asked to show their reasoning through words, but this failed to shift students towards reasoning in a Conceptual Physics frame. The frame shifts we did observe happened due to a shift at the level of a single resource. Therefore, we argue that trying to shift an individual epistemological resource may be an effective way to shift a student's epistemological frame.

Frame shifting as the result of a shift in a single epistemological resource is consistent with the findings of Rosenberg et al. In the Rosenberg et al. paper, students began completing their worksheet about the Rock Cycle by trying to accumulate as much information as possible from the external sources of information presented to them. After hearing them struggle, their teacher intervened by telling the students to ``start with what you know." Their teacher suggested the students shift their source of knowledge, and the way the students engaged with their assignment changed. Rosenberg et al. noted the underlying meaning of her statement which was that students have many ways of engaging with assignments. More importantly, Phelen, the teacher, implied that the epistemological coherence students work in can be prompted to change by a single phrase when she asked students to ``start with what [they] know."

\subsection{Future Research}

Future work should explore the ability to do the reverse and make assessment questions that prompt resource shifts.  It may be possible to create a list of recommendations for prompting resource and frame shifts in assessment writing.

Additionally, future work should explore why students often begin and stay in the Algorithmic Physics frame. It is possible that students begin in the Algorithmic Physics frame due to expectations of what `doing physics' means from before they entered the course. For example, physicists on television often discuss numbers and equations, so students may believe that they should solve problems by plugging numbers into equations. Students may begin most problems in an Algorithmic Physics frame due to their professor or textbook explicitly telling them this or from success with previous problems when using this frame. To determine which, if any, of these ideas is the case, future research should explore students' perceptions of what `doing' physics is throughout an introductory physics course. A more extensive data set across courses, professors, and student populations could also help us identify additional coherences of resources and observe different sets of epistemic frames.

\section{\label{chapter:conclusion}Conclusion}
    In this work, we explored the relationship between epistemological frames and epistemological resources during student engagement in assessment.

Students in this study typically began problems in the Algorithmic Physics frame by writing down their givens, determining what they are solving for, and searching through their equation sheet to see what they can manipulate to return a value. To those of us who have taught in introductory physics courses, this finding is not surprising. In fact, it implies the Algorithmic Frame is well-rehearsed in the students' physics courses, either in lecture, homework, assessments, or some blend of the three.

While students had a fairly stable initial frame, we found influences like intervention from authority (i.e. the interviewer) and assessment features are effective at shifting an individual epistemic resource, which shifted the students' epistemic frame. This finding suggests that although the Algorithmic Physics frame is persistent, student epistemological framing during assessment can be fluid in much the same way as it is during other classroom activities \cite{Lau_2010}.

These findings suggest a need to attend even more closely to assessment than we already do. We often think of assessment as occurring after instruction as something ``added on" at the end of instruction for us as instructors to get information about our students. This work highlights that students are also getting information from assessments. This information might impact their understandings of how they should engage in physics learning. If that is the case, then we as instructors and researchers must pay more attention to the kinds of messages we might be sending with our assessments to be sure they align with our goals for the course. Specifically, we must pay attention that we are cuing specific epistemological resources that will be productive for students. Additionally, we must be intentional about how we do that cuing so that it leads to substantive reframings. 

Further, in this work, we used the perspectives of epistemological framing and resources to make sense of student engagement in assessment. The fact that this analysis ``worked'' at all is in and of itself an important result. It was possible that the cognitive mechanisms that have been used so productively to understand learning just would not have been useful in this different context. However, we found that framing analysis is consistent with student behavior during assessments. Additionally, the fact that our findings align with other work on problem solving outside of the assessment context is also noteworthy. Again, it was possible that what students did during assessment would be entirely discontinuous with their engagement in learning. Thus, our findings that students do rely on problem-solving-like framings and their associated resources during assessment are noteworthy. Specifically, it encourages us to re-examine our own assumptions about whether and how assessment is different from other classroom activities - both for ourselves and for our students.

\section{\label{chapter:acknowledgements}Acknowledgements}
    We would like to thank Katherine Ventura and Amali Jambuge for creating and administering the tests and recording the interviews analyzed in this paper.
We would like to thank the Physics Department at Kansas State University for their support and the REU program at KSU.  This program is funded by the National Science Foundation (NSF) grant number 1757778. Any opinions, findings, and conclusions or recommendations expressed in this material are those of the author(s) and do not necessarily reflect the views of the NSF.

\bibliographystyle{apsrev}

\bibliography{main.bbl}

\end{document}